\documentclass[slac_one]{revtex4}
\usepackage{graphicx}
\usepackage{fancyhdr}
\begin{document}

\title{Elliptic flow and the high $p_T$ ridge in Au+Au collisions }

%

\author{N. K. Pruthi (for the STAR collaboration)}
\affiliation{Panjab University, Chandigarh-160014, India}
\thanks{For the full list of STAR authors and acknowledgements, see appendix `Collaborations' of this volume} 
%


\begin{abstract}
 In this paper we look for a coupling between intermediate $p_T$
  particle pairs and the $v_2$ of the remaining low $p_T$
  particles. We find that the shape of the flow vector distribution,
  which is calculated from all low $p_T$ tracks, depends in a
  non-trivial way on the angular seperation between the high $p_T$
  particles in the event. Our analysis is based on 200 GeV Au+Au
  collisions measured with the STAR
\end{abstract}

\maketitle

\thispagestyle{fancy}

\section{Introduction} 

Dihadron correlations are used in heavy-ion collisions to study the
effect of jet-quenching and other interactions of jets with the medium
created in the collision overlap zone~\cite{hardtke,fuqiang}. In
non-central collisions, this overlap zone is not symmetric. The spatial asymmetry leads to 
large azimuthal anisotropies  in momentum space resulting in high values of the second coefficient 
in the Fourier decomposition of the azimuthal distribution of produced particles with respect to reaction plane.($v_2$)~\cite{v2data}. Dihadron measurements from STAR have revealed a
correlation structure which is narrow in the azimuthal direction and
broad in longitudinal direction (the narrow $\Delta\phi$, broad
$\Delta\eta$ ``ridge'')~\cite{estructRidge,lll,putschke}. This
structure is unique to nucleus-nucleus collisions and its amplitude shows a
non-monotonic rise with increase in  collision centrality~\cite{daugherity}. We search for evidence of jet-interactions with the medium
by studying $v_2$ for events containing correlated dihadron pairs.

\section{Method and discussion}

Our analysis is based on 16 million Au+Au at 200GeV center of mass energy measured with the
STAR detector~\cite{star}. We select events with at least two tracks
having $p_T>2.0$~GeV and calculate the $|q|$ for those events. The q("flow") vector can indicate if 
particles are produced in preferential direction. $|q|$
is calculated from $q_{x}=\Sigma\cos(2\phi_i)$ and
$q_{y}=\Sigma\sin(2\phi_i)$~\cite{qdist}, where the sums run over all
tracks having $p_T<2.0$~GeV/c and $\phi_i$ is azimuthal angle of track w.r.t  reaction plane. The distribution of $|q|$ ($dN/d|q|$)
depends on the number of particles, $v_2$, and other correlations
not related to $v_2$~\cite{SorensenQM}. If there are more than two tracks
with $p_T>2.0$~GeV/c, we select the leading and sub-leading
tracks. This way we select  one dihadron pair per
event. This allows us to associate the $|q|$ of an event with the
angular seperation of a high momentum pair. In this way we can look
for modifications to $dN/d|q|$ for events with or without correlated
high $p_T$ pairs.

\begin{figure}[ht]
\begin{tabular}{lr}
\begin{minipage}{.50\textwidth}
\centering
\includegraphics[width=0.75\textwidth]{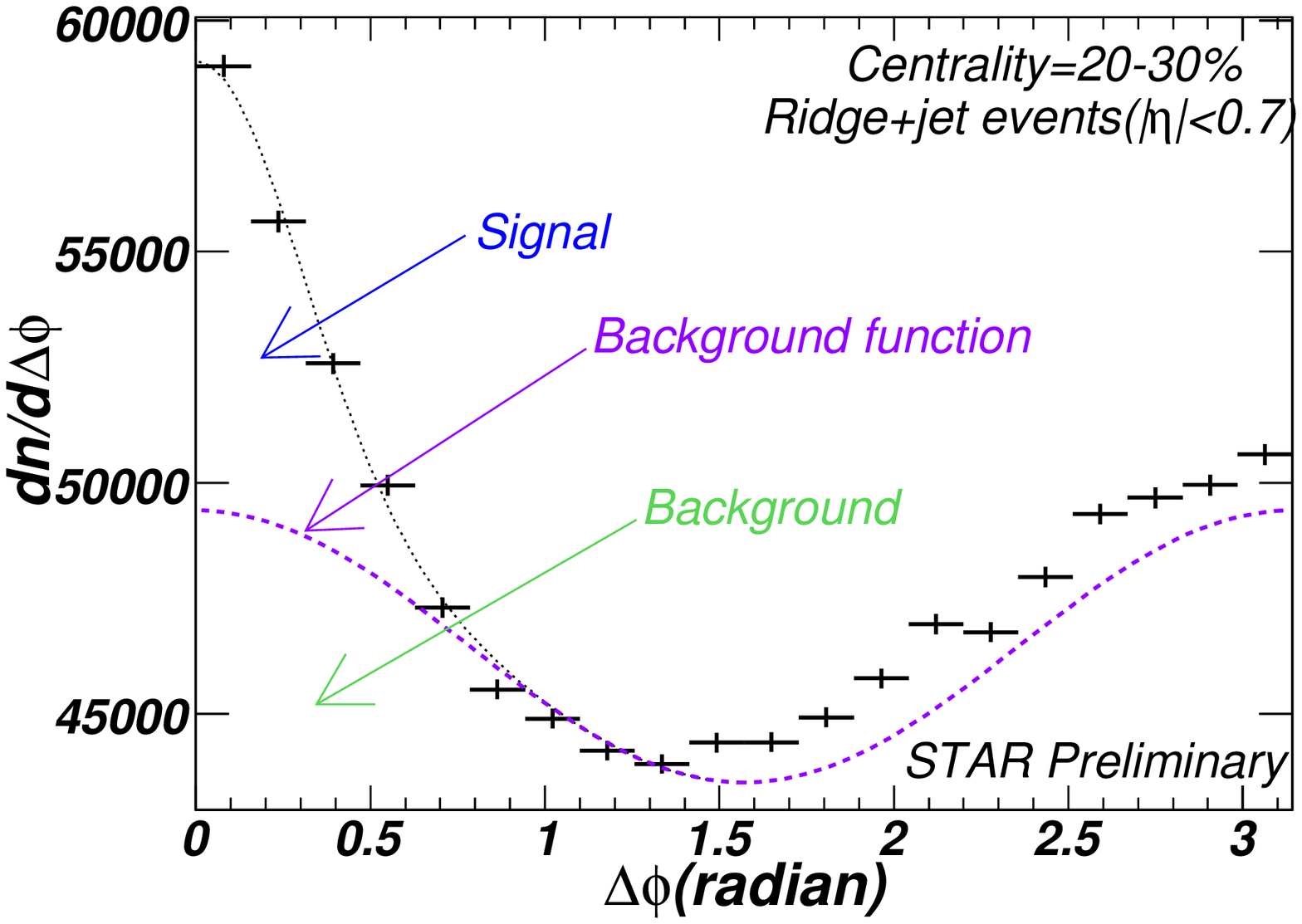}
\end{minipage} 
&\begin{minipage}{.50\textwidth}
\centering
\includegraphics[width=0.75\textwidth]{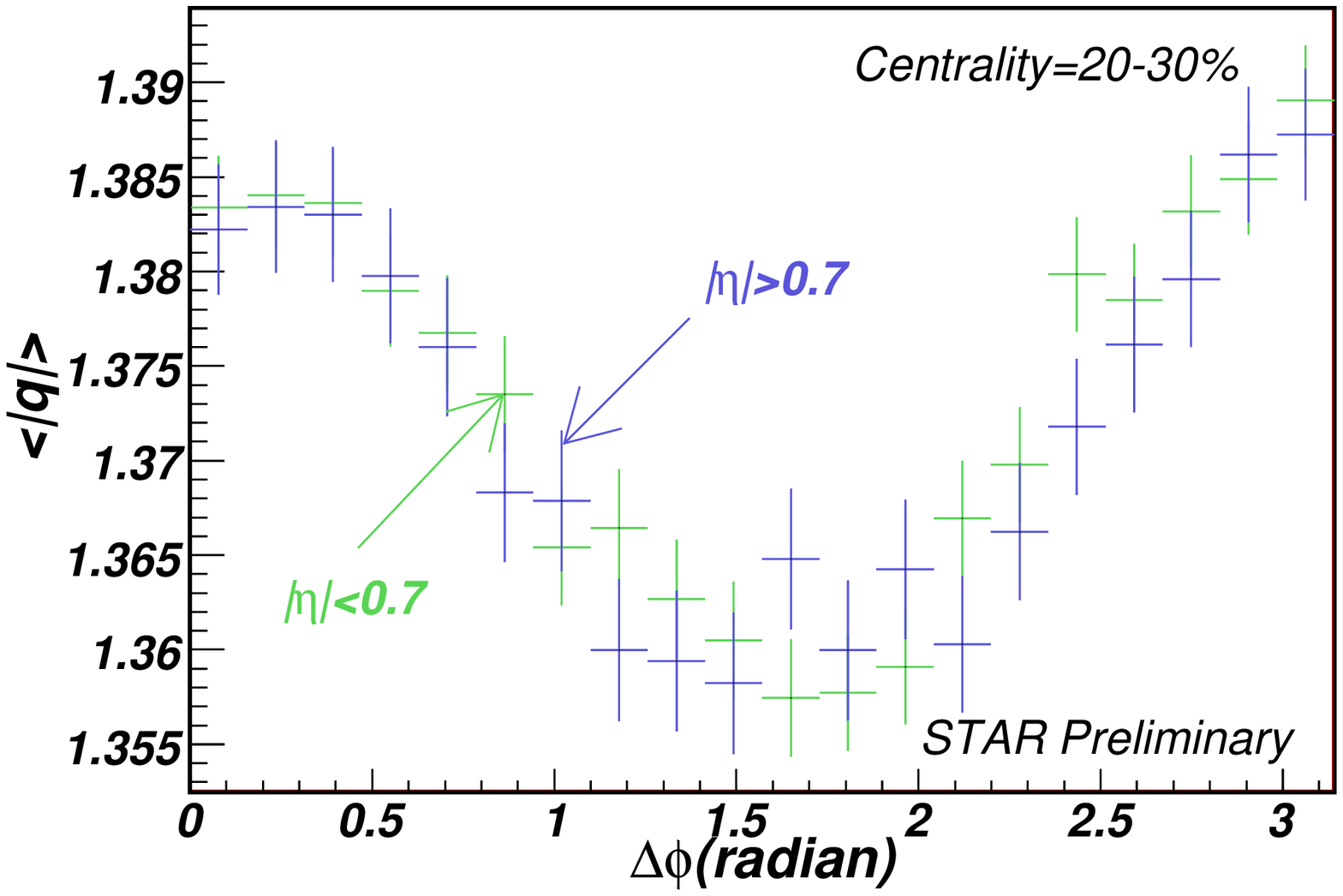}
\end{minipage} 
\end{tabular}
\caption{(left) dN/d$\Delta\phi$ for triggered events and the background
  function. (right) Non-trivial behavior of $\langle q\rangle$ with $\Delta\phi$ for ridge and ridge+jet events}
\label{fig1}
\end{figure}

Fig.~\ref{fig1} (left) shows the distribution of the relative angle between
leading and sub-leading dihadron pairs with $p_T>2.0$~GeV/c. If $v_2$
were the only source of correlations then $dN/d\Delta\phi$ should have
the following shape: $C(\Delta\phi)=b_{0}(1+2\langle
v_{2}^{trig}v_{2}^{asso}\rangle\cos(2\Delta\phi))$, where
$v_{2}^{trig}$ and $v_{2}^{asso}$ are the $v_2$ values respectively, 
for the leading and subleading (trigger and associate)
particles. Other physical processes leading to correlations are
particle decays, fragmentation of partons (jets), and HBT. Jet
correlations are expected to be prominent for high $p_T$ dihadron
pairs so a two-component model is used to fit $dN/d\Delta\phi$:
a ``jet'' component on top of a $v_2$ modulated
``background''~\cite{hardtke,fuqiang,PHdihadron}.  Assuming the two components model  is applicable, we can extract the fraction of correlated and
uncorrelated pairs for each $\Delta\phi$ bin.

The next step of our analysis is to study $dN/d|q|$ for the events
with high $p_T$ pairs as a function of the angle $\Delta\phi$ and
$\Delta\eta$ of the high $p_T$ pairs. If jets are the source of
the correlations (Signal) in Fig.~\ref{fig1} and they interact with the
medium, they may modify the $v_2$ of the event. That modification
would then change the shape of $dN/d|q|$. Or alternatively, if the
correlations beyond the $v_2$ modulation in Fig.~\ref{fig1} is caused
by the same space-momentum correlation that give rise to $v_2$, then
the correlation signal may result to large values of
$v_2$~\cite{xpcor}. This would also cause $dN/d|q|$ to depend on
$\Delta\phi$ and/or $\Delta\eta$. Fig.~\ref{fig1} (right) shows average magnitude of flow vector ($\langle
|q|\rangle$) vs. $\Delta\phi$ of the leading and subleading hadrons for
either $|\Delta\eta|>0.7$ or $|\Delta\eta|<0.7$. We observe a
non-trivial dependence on $\Delta\phi$. No prominent $\Delta\eta$
dependence is seen in this figure.

We also studied $\langle p_T\rangle$ vs. $\Delta\phi$ and find that
within errors it is independent of the angle between the high $p_T$
tracks. The multiplicity distribution for the events with two tracks
with $p_T>2.0$~GeV only deviates from the inclusive sample for
peripheral events.

Having observed a non-trivial dependence of $\langle q\rangle$ on
$\Delta\phi$, we next attempt to convert this into a $v_2$ for the
events that had a correlated pair. We categorize the events in two parts
depending on the highest $p_T$ particles (1) Ridge events :
$|(\Delta\eta)|>$ 0.7, (2) Ridge+jet events : $|(\Delta\eta)|<$0.7 and
studied the $dN/d|q|$ distributions.

\begin{figure}[ht]
\begin{tabular}{lr}
\begin{minipage}{.50\textwidth}
\centering
\includegraphics[width=0.75\textwidth]{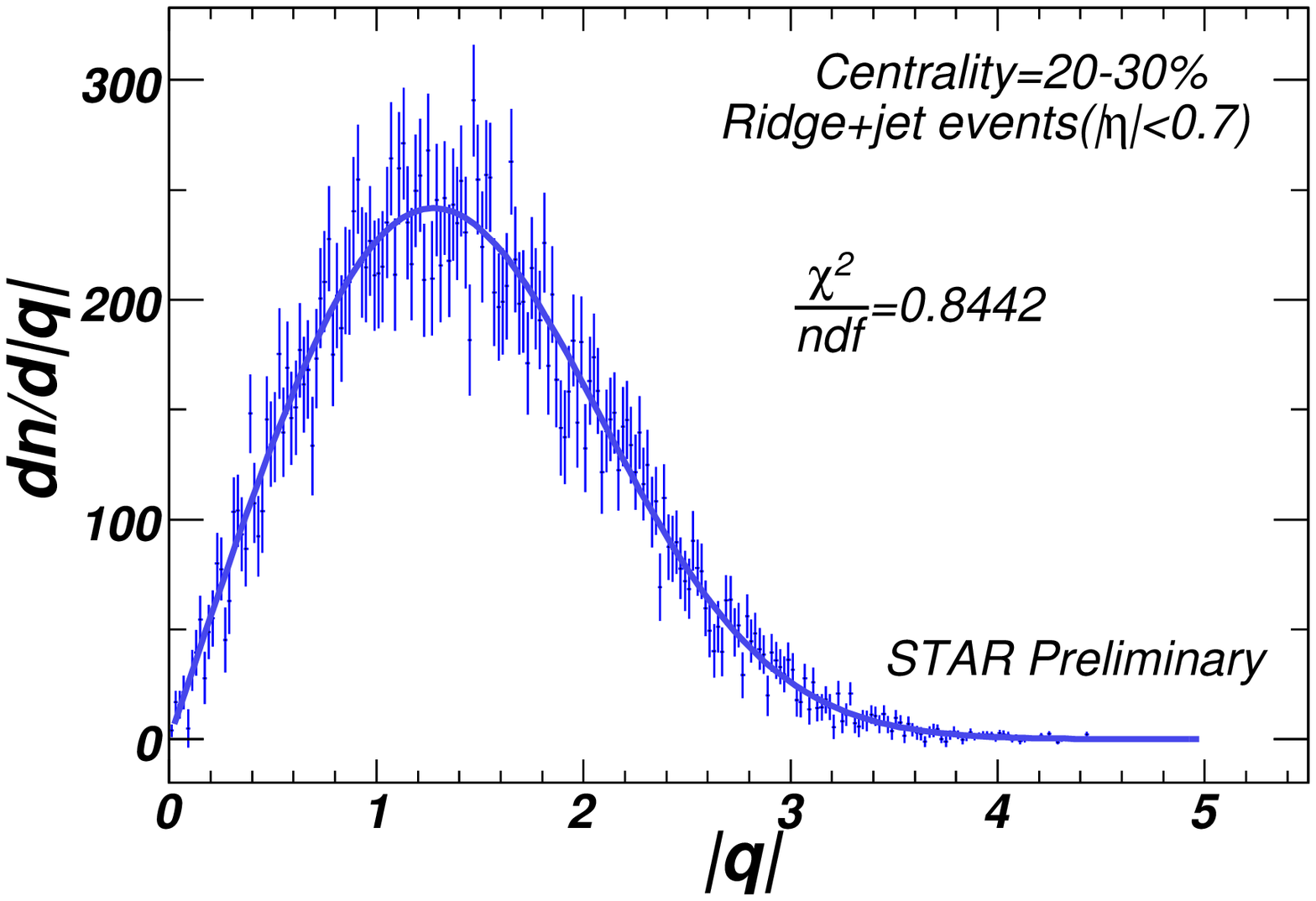}
\end{minipage} 
&\begin{minipage}{.50\textwidth}
\centering
\includegraphics[width=0.75\textwidth]{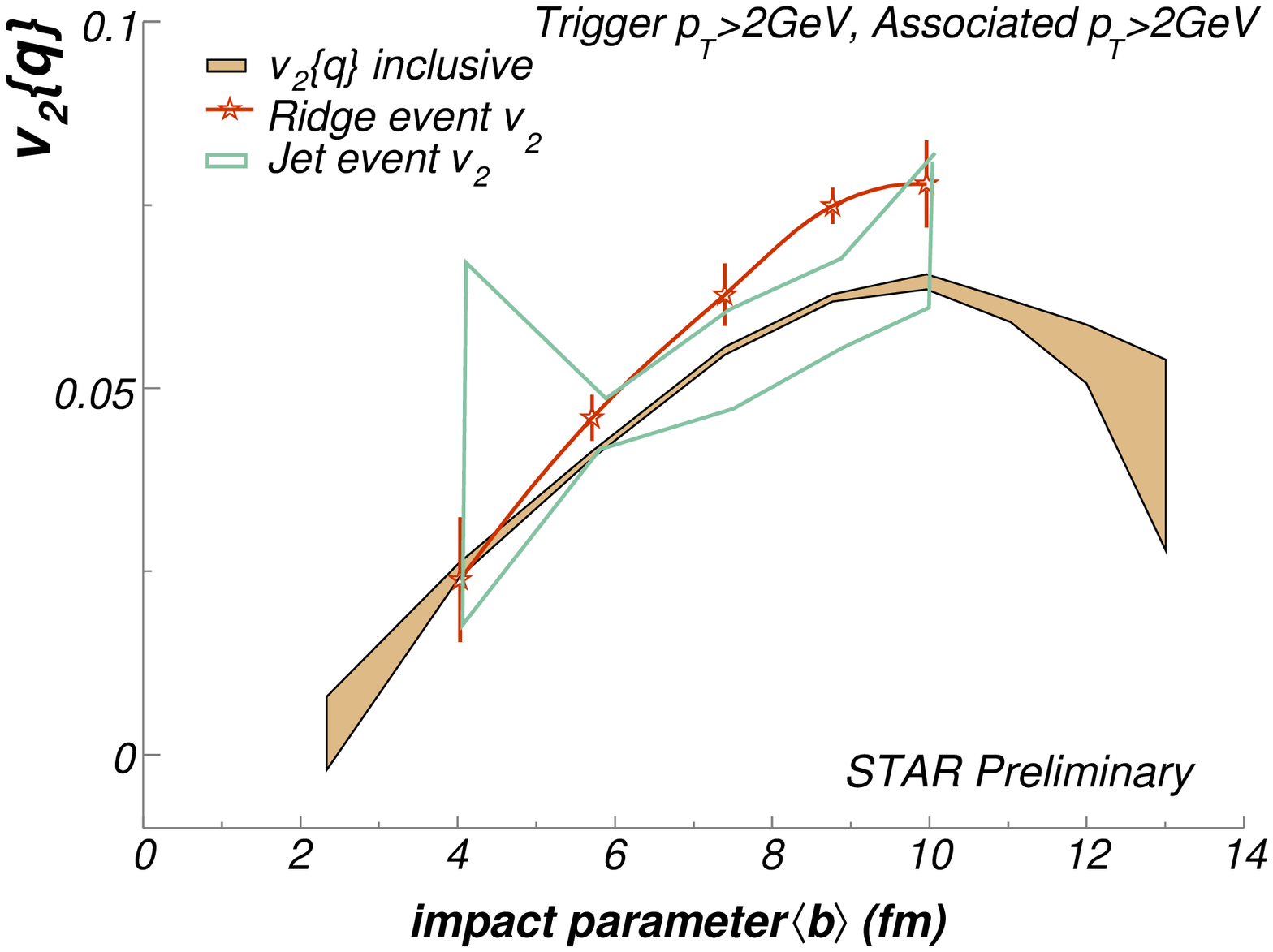}
\end{minipage} 
\end{tabular}
\caption{(left) q-vector distribution fit for events yielding correlated pair(signal)(right) Comparison of $v_{2}$ for jet signal with ridge signals }
\label{fig2}
\end{figure}

We calculate the q-distributions for the signal and background by
dividing the near-side into two bins ($\Delta\phi<4\pi/20$ and
$4\pi/20<\Delta\phi<8\pi/20$) and solving equations:
$(S_1+B_1)*dN_1=S_1*dN_S+B_1*dN_B, (S_2+B_2)*dN_1=S_2*dN_S+B_2*dN_B$.
$S_{1,2}$ is the number of events yielding a correlated pair in bin 1
or 2. $B_{1,2}$ is the number of events yielding an un-correlated pair
in bin 1 or 2. $dN_1$ and $dN_2$ are the q-distributions for events in
bin 1 and 2. $dN_S$ and $dN_B$ are the q-distributions for events
giving correlated pairs (signal) and uncorrelated pairs
(background). $dN_1$ and $dN_2$ are measured. $S_{1,2}$ and $B_{1,2}$
are extracted from the $dN/d\Delta\phi$ distributions by applying the two-component model. Then $dN_S$ and $dN_B$ can be determined
from the two equations above. Elliptic flow is calculated by fitting
the q-distributions (Fig.~\ref{fig2}(left)) for signal for ridge and
ridge+jet events. Two parameters can be extracted from the
q-distribution : $(\langle v_2\rangle^2-\sigma_{v_2}^{2}) = v_2\{q\}^2$
and $(\delta_2+2\sigma_{v_2}^{2}) = \sigma_{dyn}^2$. $\delta_2 =
\langle \cos(2(\phi_1-\phi_2))\rangle - v_2^2$.  The term 
accounts for correlations that are not related to the reaction plane and
$\sigma_{v_{2}}$ is the rms width of the $v_2$ distribution.

In Fig.~\ref{fig2}(right), we report $v_2\{2\}$ for the event classes defined
above. We use the ansatz that the near-side correlation contains a jet
part (narrow in $\Delta\phi$ and $\Delta\eta$) and a ridge part
(narrow in $\Delta\phi$ but broad in $\Delta\eta$)~\cite{putschke}.
By calculating the signal $v_{2}$ for ridge and ridge+jet events we
calculated $v_{2}$ for jet events by using equations
\begin{equation}
  Area^{ridge+jet}v_{2}^{ridge+jet}= Area^{ridge}v_{2}^{ridge}+Area^{jet}v_{2}^{jet} 
\end{equation}
\begin{equation}
  Area^{jet}= Area^{ridge+jet}-Area^{ridge}*Acceptance factor
\end{equation}
\begin{equation}
  Acceptance factor=\frac{Area^{ridge+jet (background)(\Delta\eta<0.7)}}{Area^{ridge (background)(\Delta\eta>0.7) }}
\end{equation}
This allows us to project the ridge from $|\Delta\eta|>0.7$ to
$|\Delta\eta|<0.7$. The jet is the correlation that remains.

The selection of one unique pair per event, the application  the two-component model, and the simple algebra above allows us to
calculate $v_{2}\{q\}^2=v_2^2-\sigma_{v_{2}}^2$ and
$\sigma_{q,dyn}^2=\delta_2+2\sigma_{v_2}^2$ for events giving rise to
pairs of particles in the jet-cone region, in the ridge region, and in
the background. The values of $v_{2}\{q\}$ vs. collision impact
parameter $b$ are presented in Fig.~\ref{fig2}(right). $v_{2}\{q\}$ for the events that had a high $p_T$ pair contributing to the ridge-like correlation exhibit a slightly larger
$v_2\{q\}$  for non-central and peripheral collisions than the  corresponding events contributing a pair to the background. We also attempt to determine if the $v_{2}\{q\}$ values for events
contributing pairs to the jet-like correlation are larger or smaller
than those contributing to the background. Within the current estimates  of the systematics uncertainties
the values are consistent  with $v_{2}$ from inclusive events.

\section{Summary}

We have developed an analysis to search for jet interactions with the
medium and/or possible effects linking the high $p_T$ ridge
correlation to $v_2$. The latter could occur if the same physics
mechanism underlies both phenomena as proposed in several
references~\cite{Mrowczynski:1988dz,Majumder:2007zh,Hwa:2008um,Dumitru:2008wn}. By
selecting one high-$p_T$ pair per event  and the two
component model, we find that $v_2\{q\}$ calculated from $dN/d|q|$ for
events yielding pairs in the ridge is slightly larger $v_{2}$ for non-central and peripheral collisions  than the events yielding pairs in the background.

\noindent




\end{document}